\documentclass[12pt,preprint]{aastex}

\shorttitle{Eclipsing subdwarf B binary with brown dwarf companion}
\shortauthors{Geier et al.}

\begin{document}


\title{Binaries discovered by the MUCHFUSS project\\
    SDSS\,J08205+0008 -- An eclipsing subdwarf B binary with brown dwarf companion}


\author{S. Geier, V. Schaffenroth, H. Drechsel, U. Heber,  T. Kupfer, A. Tillich, }
\affil{Dr.\,Remeis-Sternwarte, Institute for
 Astronomy, University Erlangen-N\"urnberg, Sternwartstr.~7, 96049
 Bamberg, Germany}
\email{geier@sternwarte.uni-erlangen.de}

\and

\author{R. H. \O stensen, K. Smolders, P. Degroote}
\affil{Institute of Astronomy, K.U.Leuven, Celestijnenlaan 200D, B-3001 Heverlee, Belgium}

\and

\author{P. F. L. Maxted}
\affil{Astrophysics Group, Keele University, Staffordshire, ST5 5BG, UK}

\and

\author{B. N. Barlow}
\affil{Department of Physics and Astronomy, University of North Carolina, Chapel Hill, NC 27599-3255, USA}

\and

\author{B. T. G\"ansicke, T. R. Marsh}
\affil{Department of Physics, University of Warwick, Conventry CV4 7AL, UK}

\and

\author{R. Napiwotzki}
\affil{Centre of Astrophysics Research, University of Hertfordshire, College Lane, Hatfield AL10 9AB, UK}



\begin{abstract}

Hot subdwarf B stars (sdBs) are extreme horizontal branch stars believed to originate from close binary evolution. Indeed about half of the known sdB stars are found in close binaries with periods ranging from a few hours to a few days.
The enormous mass loss required to remove the hydrogen envelope of the red-giant progenitor almost entirely can be explained by common envelope ejection. A rare subclass of these binaries are the eclipsing HW~Vir binaries where the sdB is orbited by a dwarf M star. Here we report the discovery of an HW~Vir system in the course of the MUCHFUSS project. A most likely substellar object ($\simeq0.068\,M_{\rm \odot}$) was found to orbit the hot subdwarf J08205+0008 with a period of 0.096 days. Since the eclipses are total, the system parameters are very well constrained. J08205+0008 has the lowest unambiguously measured companion mass yet found in a subdwarf B binary. This implies that the most likely substellar companion has not only survived the engulfment by the red-giant envelope, but also triggered its ejection and enabled the sdB star to form. The system provides evidence that brown dwarfs may indeed be able to significantly affect late stellar evolution. 

\end{abstract}





\section{Introduction}

Hot subdwarf B stars (sdBs) are core helium-burning stars with hydrogen envelopes too thin to sustain hydrogen shell burning and have masses of about $0.47\,M_{\rm \odot}$ \citep{heber09}. The large fraction of close binaries -- about half of the known sdB stars are members of short-period (P $\lesssim$ 10 days) binaries \citep{maxted01,napiwotzki04a} -- can be explained by binary evolution models. The required extraordinarily large mass loss in the red giant phase is triggered by the formation of a common envelope, which is finally ejected. Binary population synthesis models \citep{han02,han03} are successful in matching the observed properties of known systems qualitatively. The existence of apparently single sdB stars poses another problem. However, even in this case binary evolution comes to the rescue, because such stars may form from the merger of two helium white dwarfs \citep{webbink84,ibentutukov84} or from the engulfment and possible destruction of a substellar object \citep{soker98,nelemans98}.

The existence of eclipsing sdB+dM binaries of HW\,Vir type with very short orbital periods ($0.1-0.26\,{\rm d}$) and very low companion masses between $0.1\,M_{\rm \odot}$ and $0.2\,M_{\rm \odot}$ \citep[e.g.][]{for10, oestensen10} shows that stars close to the nuclear burning limit of $\simeq0.08\,M_{\rm \odot}$ are able to eject a common envelope and form an sdB. Claims have been made that the primaries of AA~Dor \citep{rauch00,rucinski09} and HS~2231$+$2441 \citep{oestensen08} are low mass sdBs and the companions therefore substellar. However, in the former case this has been refuted by the measurement of the companion's radial velocity (RV) curve \citep{vuckovic08} and in the latter case by the gravity measurement \citep{for10}. Substellar companions to sdB stars have been found using the light travel time technique \citep[][and references therein]{schuh10}. However, these systems have wide orbits and are therefore unlikely to have experienced a common envelope phase. None of these companions influenced the evolution of its host star, but the presence of such objects in wide orbits may be an indication of current or former close substellar companions to sdBs.

Here we report the discovery of the short-period eclipsing HW~Vir type binary J08205+0008 from the MUCHFUSS project. This system is regarded as the first one in which a close substellar companion to an sdB star has been detected unambiguously. 

\section{Target selection and radial velocity curve}\label{rv}

The project Massive Unseen Companions to Hot Faint Underluminous Stars from SDSS (MUCHFUSS) aims at finding sdBs with compact companions like massive white dwarfs (WDs, $M>1.0\,M_{\rm \odot}$), neutron stars or black holes. Details on the survey and target selection procedure are provided in \citet{geier10}, and an analysis of seven sdB binaries in \citet{geier11}. The same selection criteria that we applied to find such binaries are also well suited to single out hot subdwarf stars with constant high radial velocities in the Galactic halo and search for hypervelocity stars. First results of the search for hypervelocity stars are presented in \citet{tillich10}. 

The MUCHFUSS target selection strategy is tailored to single out RV variations on time scales of half an hour or less. Such variations may indicate the presence of short-period systems of relatively low RV amplitude or longer-period binaries with high RV amplitudes. The latter are the prime targets for the core programme of MUCHFUSS. Obviously, the campaign is also bound to find short-period, low RV amplitude systems with low mass stellar or even substellar companions.

SDSS\,J082053.53+000843.4 (GALEX\,J082053.6+000843, in short J08205+0008, $g=14.9\,{\rm mag}$) was classified as an sdB star by colour selection and visual inspection of SDSS spectra \citep{abazajian09}, which are flux calibrated and cover the wavelength range from $3800\,{\rm \AA}$ to $9200\,{\rm \AA}$ with a resolution of $R=1800$. The six individual sub-spectra showed significant RV variability and the star became a high-priority target for the MUCHFUSS spectroscopic follow-up. Eighteen spectra were taken with the EFOSC2 spectrograph ($R\simeq2200,\lambda=4450-5110\,{\rm \AA}$) mounted at the ESO\,NTT. Five additional spectra were taken with the Goodman spectrograph mounted at the SOAR telescope ($R\simeq2500, \lambda=3500-6160\,{\rm \AA}$). 

\begin{figure}[t!]
\begin{center}
	\resizebox{16cm}{!}{\includegraphics{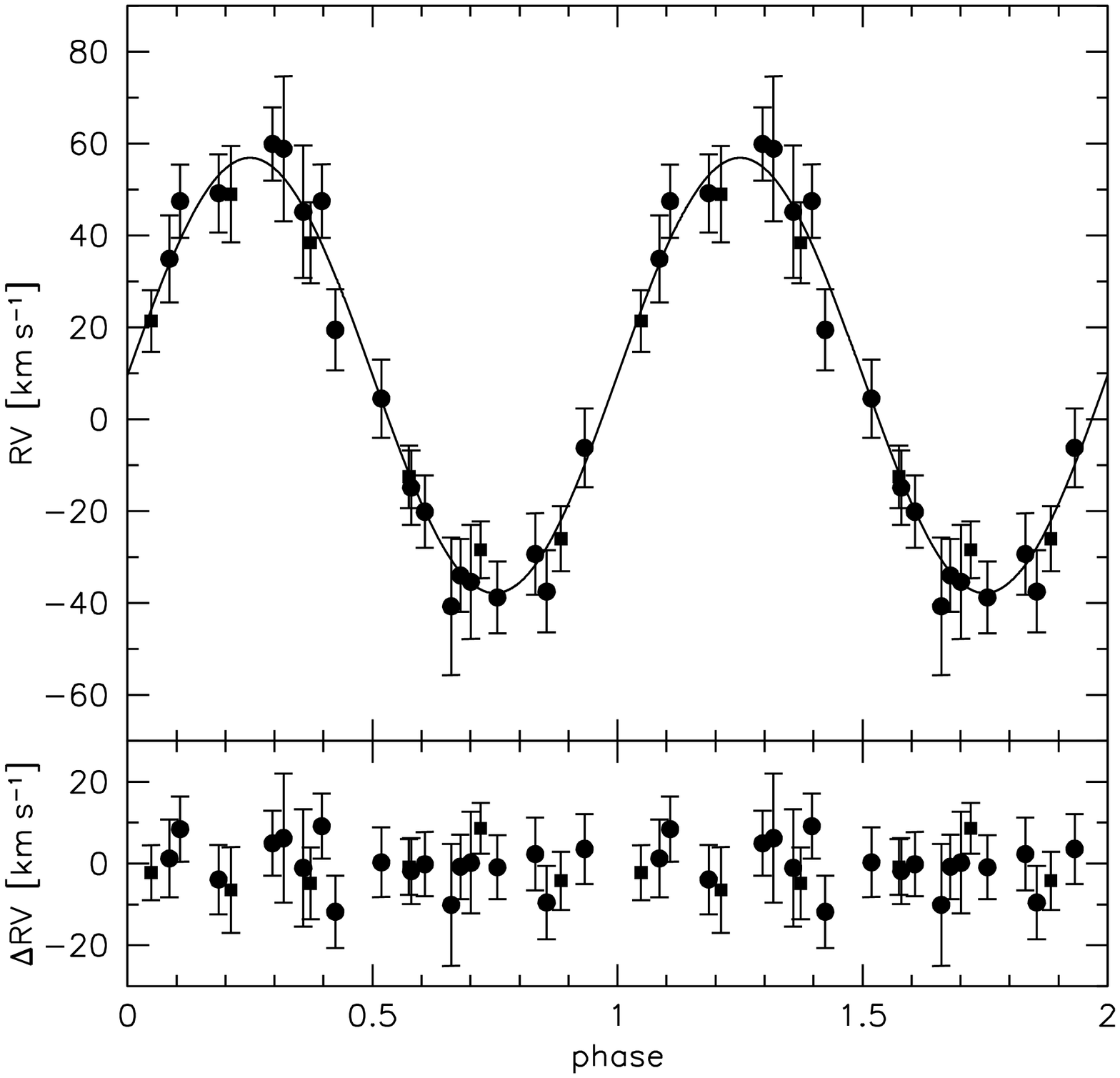}}
\end{center}
\caption{Radial velocity plotted against orbital phase of J08205+0008. The RV data were phase folded with the most likely orbital period. The residuals are plotted below. The RVs were measured from spectra obtained with SDSS (rectangles) and ESO-NTT/EFOSC2 (circles). The errors are formal $1\sigma$ uncertainties.}
\label{RV_J082053}
\end{figure}

The radial velocities were measured by fitting a set of mathematical functions to the hydrogen Balmer lines as well as He\,{\sc i} lines using the FITSB2 routine \citep{napiwotzki04b}. The continuum is matched by a polynomial, line wings and line core by a Lorentzian and a Gaussian, respectively. The orbital solution (see Table~\ref{tab:par}) was derived based on the spectra from SDSS and the ones taken with EFOSC2 as described in \citet{geier11}. The phase coverage of the RV curve is very good (see Fig.~\ref{RV_J082053}). The best fit orbital period is $0.096\pm0.001\,{\rm d}$. Two alias periods are possible at P=$0.088\,{\rm d}$ and $0.108\,{\rm d}$, but can be rejected by the analysis of the light curve. 

\section{The light curve}

\begin{figure}[t!]
\begin{center}
	\resizebox{16cm}{!}{\includegraphics{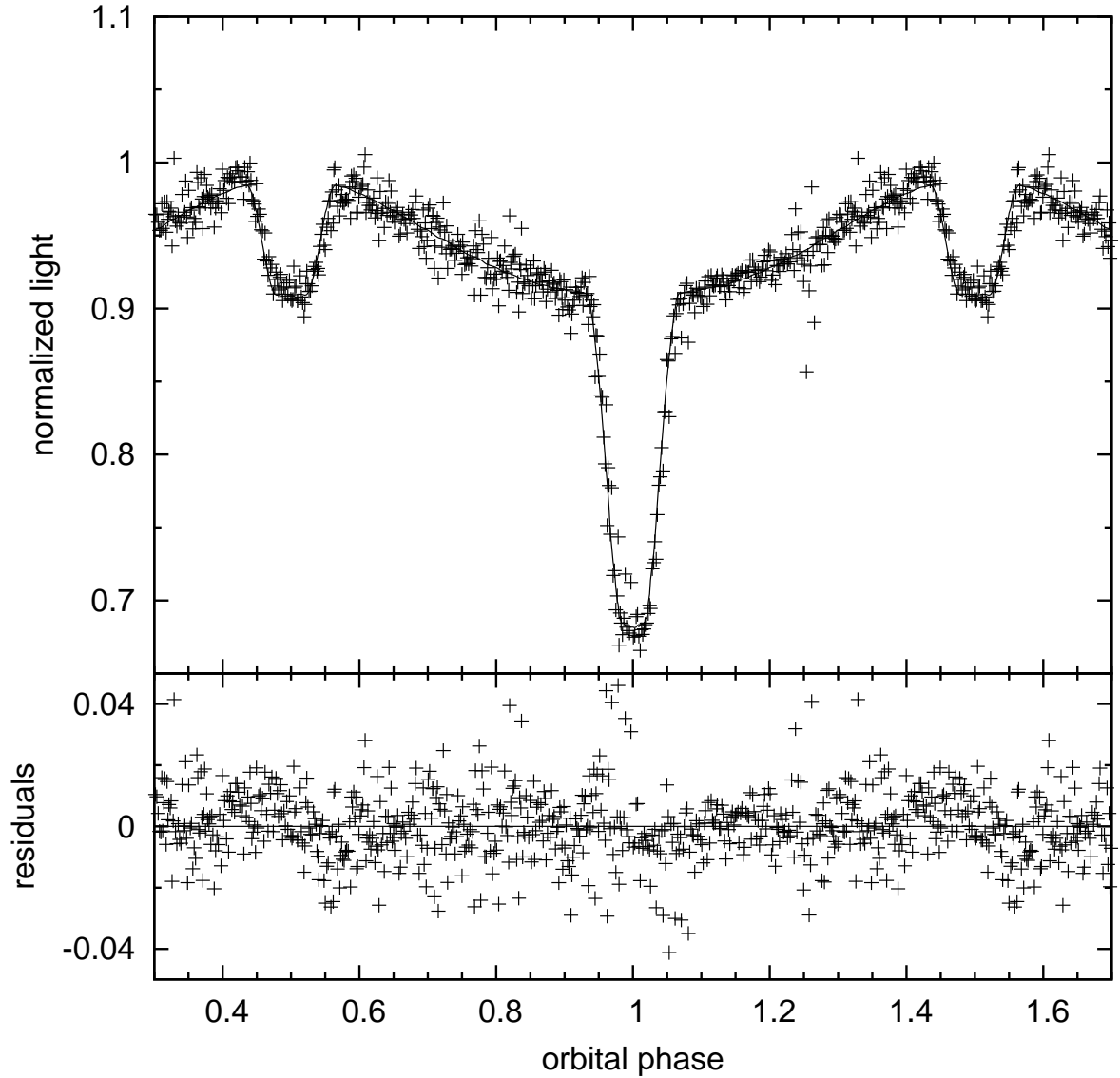}}
\end{center}
\caption{Phased R-band light curve of J08205+0008. The slight asymetry in the secondary eclipse is most likely an artifact caused by the combination of the different datasets. The model light curve for the post-RGB case is overplotted as solid line. Fits of similar quality can be obtained for different mass ratios.}
\label{light}
\end{figure}

Photometric light curves were obtained on three separate nights between November 29, 2009 and January 13, 2010, with the Flemish 1.2-m Mercator Telescope on La Palma, Canary Islands. During the period that these observations were made, we were in the process of commissioning an upgrade to the Merope CCD camera, fitting it with a large E2V frame transfer CCD with 2048$\times$3074 pixels \citep{oestensen10a}. Although the upgraded camera suffered from some technical issues, and the observing conditions were far from perfect, the deep primary and shallow secondary eclipses superimposed on a strong reflection effect are perfectly clear in all the light curves (see Fig.~\ref{light}). The data were obtained using the Geneva R-band filter and exposure times from 16 to 90 seconds.

The light curve was analysed with the MORO code, which is based on the Wilson-Devinney approach \citep{wilson71} but takes into account radiative interaction between the components of hot, close binaries \citep{drechsel95}. 

The light curve was binned over narrow time intervals. The photometric period was determined by measuring the timespans between three consecutive primary eclipses. The result $0.096\pm0.001\,{\rm d}$ is perfectly consistent with the best spectroscopic solution (see Sect.~\ref{rv}). We used Wilson-Devinney mode 2, which poses no restrictions to the system configuration and links the luminosity and the temperature of the second component by means of the Planck law. The gravity darkening exponents ($g_{1,2}$) and the linear limb darkening coefficient of the sdB primary ($x_1(R)$) were fixed at literature values \citep[][and references therein]{drechsel01}, whereas the linear limb darkening coefficient of the secondary ($x_2(R)$) was fixed to $1.0$ because all light curve solutions converged at this value. The bolometric albedo of the primary was also fixed to $1.0$. The temperature of the sdB was taken from the spectral analysis ($T_{\rm eff}(1)=26\,000\,{\rm K}$). 

The remaining adjustable parameters are the inclination, the temperature of the second component ($T_{\rm eff}(2)$), the Roche potentials ($\Omega_{1,2}$), the bolometric albedo of the secondary ($A_{2}$), the radiation pressure parameter ($\delta_1$) and the luminosity of the hot component. The fractional Roche radii ($r_{1,2}$) in units of the orbital separation $a$ were calculated using the Roche potentials and the mass ratio $q$. We used the binary mass function derived from spectroscopy to calculate possible mass ratios for a range of primary masses. A grid of light curve solutions with different mass ratios was calculated. In order to derive errors we created $500$ new datasets with a bootstrapping algorithm by random sampling with replacement from the original dataset. In each case a light curve solution was calculated in the way described above. The standard deviations of these results were adopted as the error estimates for the parameters. 

The flat-bottomed eclipses, and the way the secondary eclipses reach down to the flux level just before and after primary eclipse, indicate that the secondary is totally eclipsed by the primary. The eclipses ensure that the inclination of the system and the relative radii of the components are very well constrained, but the usual degeneracy still remains in the mass ratio and the ratio of the components' radii and the size of the orbit. Table~\ref{tab:par} shows light curve solutions for the most likely sdB masses after combining the photometric and spectroscopic analyses. Fig.~\ref{light} shows an example model fit to the light curve. 

\section{Atmospheric and stellar parameters of the sdB star \label{atmo}}

Atmospheric parameters have been determined by fitting a grid of synthetic spectra, calculated from line-blanketed, solar-metalicity LTE model atmospheres \citep{heber00}, to the hydrogen Balmer and helium lines of the SDSS and SOAR spectra in the way described in \citet{geier10}. The single spectra have been corrected for their orbital motion and coadded. In order to investigate systematic effects introduced by the individual instruments, especially the different resolutions and wavelength coverages, the parameters have been derived separately from spectra taken with SDSS ($S/N=83$, $T_{\rm eff}=26000\pm1000\,{\rm K}$, $\log{g}=5.37\pm0.14$) and SOAR ($S/N=61$, $T_{\rm eff}=26900\pm300\,{\rm K}$, $\log{g}=5.51\pm0.04$), respectively. The weighted mean values have been calculated and adopted as final solutions (see Table~\ref{tab:par}). 

The contribution of light from the irradiated surface of the cool companion in HW\,Vir type binaries can lead to systematic shifts in the atmospheric parameters \citep[e.g.][]{heber04,for10,mueller10}. The quality of the individual SDSS spectra, which cover most of the orbital phase is not good enough to resolve this effect. The statistical errors are higher than the expected modulations. \citet{drechsel01} adopted errors of $900\,{\rm K}$ in $T_{\rm eff}$ and $0.1\,{\rm dex}$ in $\log{g}$ to account for this effect in the case of the HW\,Vir type binary HS\,0705$+$6700. Since the orbital period as well as the atmospheric parameters of this binary are similar to the ones of J08205+0008, we adopt similar uncertainties here.  Systematic errors introduced by different model grids are typically smaller than that \citep{lisker05,geier07}. 

While the inclination of the system is well constrained by the light curve analysis, there remains a degeneracy between the masses and the radii of the components. The surface gravity of the sdB determined in the quantitative spectroscopic analysis provides an additional constraint since it only depends on the mass and the radius of the subdwarf. To proceed, however, we need to constrain the sdB mass from evolutionary models.

\begin{figure}[t!]
\begin{center}
	\resizebox{16cm}{!}{\includegraphics{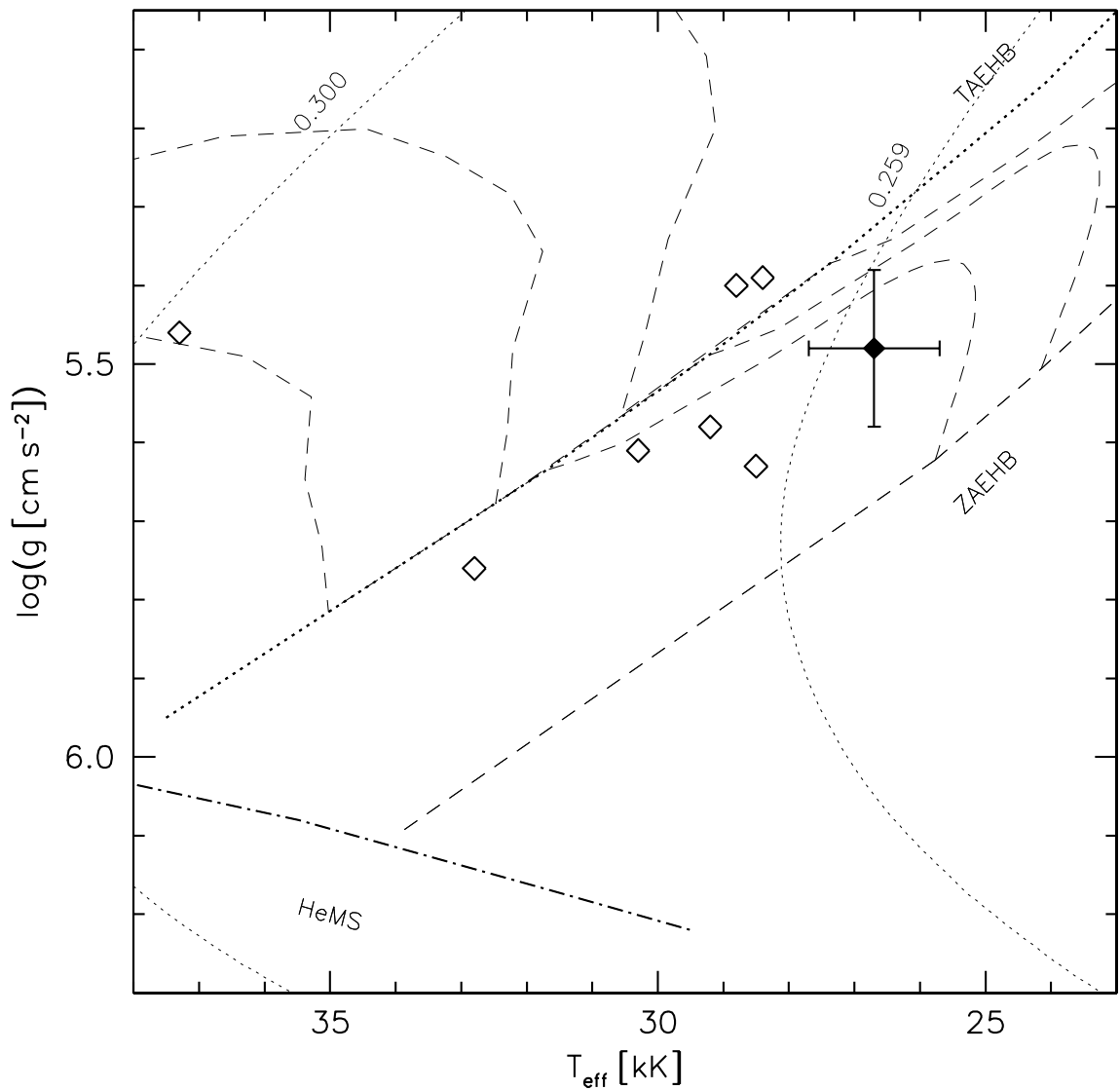}}
\end{center}
\caption{$T_{\rm eff}-\log{g}$-diagram. The helium main sequence (HeMS) and the EHB band (limited by the zero-age EHB, ZAEHB, and the terminal-age EHB, TAEHB) are superimposed with EHB evolutionary tracks from \citet{dorman93} and a post-RGB track from \citet{driebe98}. The position of J08205+0008 is indicated with a solid diamond. Open diamonds mark the position of other HW\,Vir-like systems \citep{charpinet08, drechsel01, for10, maxted02, mueller10, oestensen08, wood99}.}
\label{tefflogg}
\end{figure}

In Fig.~\ref{tefflogg} we compare the position of the star in the 
($T_{\rm eff}$, $\log{g}$)-plane to other HW~Vir stars as well as to two sets of models. The first one represents the canonical picture of EHB evolution, while the second one recalls post-RGB evolution, which means that the sdB star has left the RGB early and did not ignite helium in the core at all. 

According to Fig.~\ref{tefflogg} the star is situated on the Extreme Horizontal Branch (EHB) consistent with being a core helium-burning star as are the other HW~Vir stars (except AA~Dor). Since the orbital period is short, it was formed via common envelope ejection. Population synthesis models \citep{han02,han03} predict a mass range of $M_{\rm sdB}=0.37-0.48\,M_{\rm \odot}$ with a sharp peak at $0.47\,M_{\rm \odot}$ for sdBs in binaries formed in this way. Even lower sdB masses (down to $0.3\,M_{\rm \odot}$) are possible, when a more massive progenitor star ($2-3\,M_{\rm \odot}$) ignites core helium-burning under non-degenerate conditions. Since this formation channel is predicted to be rare, the results of the quantitative spectroscopic analysis are fully consistent with an EHB model of the canonical mass. Hot subdwarf masses derived from asteroseismic analyses \citep[e.g.][]{vangrootel10} or from analyses of eclipsing binaries are in general agreement with this picture. 

In spite of the consistency discussed above, it may still be premature to adopt the canonical mass. It has been pointed out that the sdB stars in AA\,Dor, HS\,2333$+$3927 and HD\,188112  might not burn helium in their cores and may therefore be of lower mass \citep{rauch00,heber04,heber03}. Such close binaries are expected to form whenever the RGB evolution is interrupted by the ejection of a common envelope before the core has reached the mass required for helium ignition.

\citet{driebe98} calculated evolutionary tracks of these so-called post-RGB objects. In Fig.~\ref{tefflogg} we see that the evolutionary track for a $\simeq0.25\,M_{\rm \odot}$ post-RGB star is also consistent with the atmospheric parameters of J08205+0008. The time it takes for such a low mass star to cross the region of the $T_{\rm eff}-\log{g}$ plane where canonical subdwarfs spend their time is only $\simeq2\,{\rm Myr}$ (for the model plotted in Fig.~\ref{tefflogg}), compared to $100-150\,{\rm Myr}$ for canonical EHB stars. Hence, such a low mass solution is unlikely, but cannot be completely ruled out. In Table~\ref{tab:par} we show solutions for both the canonical EHB and the post-RGB scenario.

To further constrain the mass of the sdB star we can make use of the spectroscopic gravity determination. However, the $\log{g}$ derived from the spectra is fully consistent with both solutions within the uncertainties. To tighten this constraint, the gravity would need to be determined to better precision ($\Delta\log{g}\simeq0.05$). 

\section{Constraining the nature of the companion \label{comp}}

\begin{figure}[t!]
\begin{center}
	\resizebox{16cm}{!}{\includegraphics{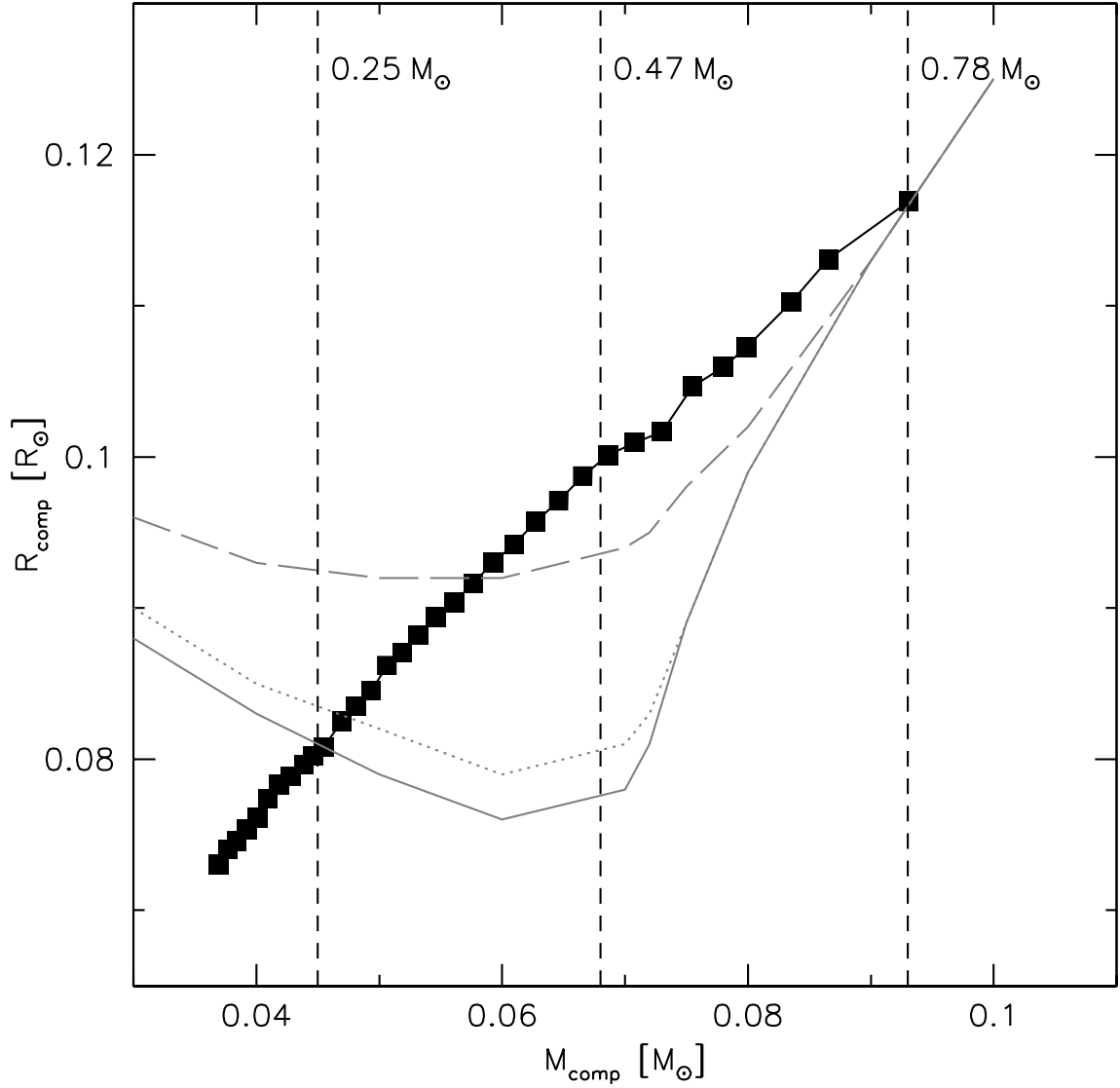}}
\end{center}
\caption{The radius of the cool companion plotted against the mass. The black symbols mark the relation derived from the spectroscopic and photometric analysis. The three curves mark theoretical relations for substellar objects with of $1$ (dashed line), $5$ (dotted line) and $10\,{\rm Gyr}$ (solid line) taken from \citet{baraffe03}. The dashed vertical lines mark different values of the corresponding sdB masses.}
\label{mr-relation}
\end{figure}

We now turn to the cool companion. Fig.~\ref{mr-relation} shows its mass-radius relation derived from the  
light curve analysis. Theoretical mass-radius relations for low mass stellar and substellar objects taken from \citet{baraffe03} are given for comparison. Since the normal progenitors of sdBs are expected to be stars around $1.0\,M_{\rm \odot}$, it seems reasonable to assume an age of a few Gyr for the system. The relations intersect at two very different mass ranges -- a low mass one close to $0.25\,M_{\rm \odot}$ and a high mass one at $0.78\,M_{\rm \odot}$. The high mass solution is very unlikely, because such high sdB masses are neither predicted by theory \citep{han02, han03, zhang09} nor have ever been measured empirically \citep[e.g. ][]{vangrootel10}.

For the canonical sdB mass of $0.47\,M_{\rm \odot}$ and the corresponding companion mass of $0.068\,M_{\rm \odot}$, the radius of the companion would have to be $\simeq20\%$ larger than expected from theory (see Fig.~\ref{mr-relation}). The irradiation by the sdB primary should influence the radius of the companion, which may be inflated. This effect has been measured in the case of hot Jupiter planets \citep[e.g.][]{udalski08}, the accreting WD+BD binary SDSS\,103533.03$+$055158.4 \citep{littlefair06} and the eclipsing MS+BD binary CoRoT-15b \citep{bouchy11}. \citet{baraffe03} calculated mass-radius relations for planets including the influence of irradiation. The effect is of the order of up to $10\%$ for hot Jupiters orbiting solar type stars. More massive objects should be less affected. On the other hand the sdB is more luminous than a solar type star and the separation of the system is less than one solar radius. Recently \citet{parsons10} modelled the inflation of a late M star irradiated by a hot WD ($T_{\rm eff}=57\,000\,{\rm K}$) in a $0.13$ day orbit and showed that the measured $10\%$ increase in radius compared to theoretical models can be reproduced in this way. For a discussion of other possible solutions to this issue (e.g. cold spots on the BD surface) see \citet{bouchy11}. 

For a low sdB mass of $\simeq0.25\,M_{\rm \odot}$ the companion's mass-radius relation is also consistent with theory. In this case the companion mass would be $0.045\,M_{\rm \odot}$. A third possibility may be that the progenitor of the sdB was originally more massive. In this case the sdB mass could be as low as $\simeq0.3\,M_{\rm \odot}$ and the system as well as the substellar companion would be much younger. Since young BDs are considerably larger, this may also lead to a consistent solution (see Fig.~\ref{mr-relation}).

In the cases discussed above, the mass of the companion ranges between $0.045_{-0.002}^{+0.003}\,M_{\rm \odot}$ and $0.068_{-0.003}^{+0.003}\,M_{\rm \odot}$. The most conservative theoretical lower limit for core hydrogen-burning \citep[$\simeq0.07\,M_{\rm \odot}$,][]{chabrier00} is right at the border of this range. We therefore conclude that the companion is most likely a brown dwarf. However, given that the sdB mass is not strictly constrained, the companion may also be a star of extremely low mass.

\section{Discussion}

We have presented a spectroscopic and photometric analysis of the HW~Vir type eclipsing sdB star J08205+0008, discovered in the course of the MUCHFUSS project. {The companion turns out to be a very low-mass, most likely substellar object.}  

Although the mass of the sdB is not yet tightly constrained, it is important to stress that the companion remains below the core hydrogen-burning limit for reasonable subdwarf masses ranging from $0.25\,M_{\rm \odot}$ to $0.47\,M_{\rm \odot}$. The inclination constraint from eclipses means that J08205+0008 has the lowest unambiguously measured companion mass yet found in a subdwarf B binary.

The question whether the sdB is burning helium in its core or not remains open for now. Time resolved high resolution spectroscopy is necessary to measure both the $v_{\rm rot}\sin{i}$ and the $\log{g}$ of the subdwarf with high accuracy. Combined with a high-quality multi-colour light curve much tighter constraints could be put on this unique binary system. The fact that the sdB is situated on the EHB is a strong argument in favour of the EHB-scenario, because post-RGB objects are very rare and not related to the EHB. 

As witnessed by the HW\,Vir type systems, stellar companions with masses as low as $0.1\,M_{\rm \odot}$ are able to eject a common envelope and form an sdB star without being destroyed. The case of J08205+0008 demonstrates that even lower mass objects, i.e. substellar objects, are sufficient. This finding can be used to constrain theoretical models \citep{soker98,nelemans98} and learn more about the role of substellar companions for the formation of single and close binary sdBs.  

The double-lined spectroscopic WD+BD system WD\,0137$-$349 \citep{maxted06} is a binary very similar to J0820+0008, but in a later stage of evolution. It consists of a He-core white dwarf of $0.39\,M_{\rm \odot}$ orbited by a $0.053\,M_{\rm \odot}$ brown dwarf in 0.0803 days. When evolving on the white dwarf cooling sequence J08205+0008 will therefore appear as a twin to WD\,0137$-$349 once it is cooled down to the effective temperature of the latter ($15\,000\,{\rm K}$).




\acknowledgments

Based on observations at the La Silla Observatory of the 
European Southern Observatory for programmes number 082.D-0649 and 084.D-0348 and on observations with the Southern Astrophysical Research (SOAR) telescope operated by the U.S. National Optical Astronomy Observatory (NOAO), the Ministério da Ciencia e Tecnologia of the Federal Republic of Brazil (MCT), the University of North Carolina at Chapel Hill (UNC), and Michigan State University (MSU). Based on observations collected with the Flemish 1.2-m Mercator Telescope at the Roque de los Muchachos, La Palma, Spain.
A.T. and S.G. are supported by the Deutsche Forschungsgemeinschaft (DFG) through grants HE1356/45-1 and HE1356/49-1. R.\O. acknowledges funding from the European Research Council under the European Community's Seventh Framework Programme (FP7/2007--2013)/ERC grant agreement N$^{\underline{\mathrm o}}$\,227224 ({\sc prosperity}), as well as from the Research Council of K.U.Leuven grant agreement GOA/2008/04.

\clearpage



\clearpage

\clearpage

\clearpage

\clearpage


\clearpage

\clearpage

\begin{table}[h]
\caption{Parameters of J08205+0008 \label{tab:par}}
\vspace{0.5cm}
\begin{tabular}{lccc}
\noalign{\smallskip}
\multicolumn{4}{l}{SPECTROSCOPIC PARAMETERS}\\
\noalign{\smallskip}
\hline
\noalign{\smallskip}
$T_{\rm 0}$&[HJD]&\multicolumn{2}{c}{$2455147.8564\pm0.0006$} \\
$P$&[d]&\multicolumn{2}{c}{$0.096\pm0.001$} \\
$\gamma$& [${\rm km\,s^{-1}}$] & \multicolumn{2}{c}{$9.5\pm1.3$} \\
$K_{\rm 1}$& [${\rm km\,s^{-1}}$] & \multicolumn{2}{c}{$47.4\pm1.9$} \\
$f(M)$& [$M_{\rm \odot}$] & \multicolumn{2}{c}{$0.0011\pm0.0001$} \\
$T_{\rm eff}$&[K]&\multicolumn{2}{c}{$26700\pm1000$} \\
$\log{g}$&& \multicolumn{2}{c}{$5.48\pm0.10$} \\
$\log{y}$&& \multicolumn{2}{c}{$-2.0\pm0.07$} \\
$M_{\rm 2, M_{\rm sdB}=0.47}$ & [$M_{\rm \odot}$] & \multicolumn{2}{c}{$0.068_{-0.003}^{+0.003}$} \\
$M_{\rm 2, M_{\rm sdB}=0.25}$ & [$M_{\rm \odot}$] & \multicolumn{2}{c}{$0.045_{-0.002}^{+0.003}$} \\
\noalign{\smallskip}
\hline
\noalign{\smallskip}
\multicolumn{4}{l}{LIGHT CURVE SOLUTION}\\
\noalign{\smallskip}
\hline
\noalign{\smallskip}
$A_1$& &\multicolumn{2}{c}{1.0}\\
$T_{\rm eff}(1)$&[K]&\multicolumn{2}{c}{26700}\\
$g_1$& &\multicolumn{2}{c}{1.0}\\
$g_2$& &\multicolumn{2}{c}{0.32}\\
$x_1(R)$& &\multicolumn{2}{c}{0.18}\\
$x_2(R)$&&\multicolumn{2}{c}{1.0}\\
&&$M_{\rm sdB}=0.25\,M_{\rm \odot}$&$M_{\rm sdB}=0.47\,M_{\rm \odot}$\\
\cline{3-4}
\noalign{\smallskip}
$q\,(=M_{2}/M_{1})$ & & $0.181$ & $0.1438$\\
\noalign{\smallskip}
\hline
\noalign{\smallskip}
$i$ & [$^{\rm \circ}$] & $85.87\pm0.16$ & $85.83\pm0.19$ \\
$q\,(=M_{2}/M_{1})$ & & $0.181$ & $0.1438$\\
$T_{\rm eff}(2)$ & [K]& $2958\pm207$ & $2484\pm230$\\
$A_2$ & & $1.11\pm0.05$ & $1.09\pm0.04$\\
$\Omega_1$&&$3.687\pm 0.026$&$3.621\pm 0.027$\\
$\Omega_2$&&$2.732\pm 0.009$&$2.470\pm 0.007$\\
$\frac{L_1}{L_1+L_2}(R)$&&$0.99983\pm 0.00008$&$0.99995\pm 0.00004$\\
$\delta_1$&&$0.0282\pm 0.002$&$0.030\pm 0.002$\\
$r_1$(pole)&[a]&$0.277 \pm 0.002$&$0.278\pm 0.002$\\
$r_1$(point)&[a]&$0.283\pm 0.002$&$0.284\pm 0.002$\\
$r_1$(side)&[a]&$0.280\pm 0.002$&$0.282 \pm 0.002$\\
$r_1$(back)&[a]&$0.282\pm 0.002$&$0.283 \pm 0.002$\\
\noalign{\smallskip}
$r_2$(pole)&[a]&$0.136 \pm 0.001$&$0.137\pm 0.001  $\\
$r_2$(point)&[a]&$0.137 \pm 0.001$&$0.138 \pm 0.001$\\
$r_2$(side)&[a]&$0.137\pm 0.001 $&$0.138 \pm 0.001$\\
$r_2$(back)&[a]&$0.140 \pm 0.001$&$0.142 \pm 0.001 $\\
\noalign{\smallskip}
\hline
\noalign{\smallskip}
\end{tabular}\\ 
\end{table}




\end{document}